\documentstyle[twocolumn,prl,aps]{revtex}
\begin{document}
\twocolumn[\hsize\textwidth\columnwidth\hsize\csname @twocolumnfalse\endcsname
\draft
\def \beq{\begin{equation}}
\def \eeq{\end{equation}}
\def \beqarr{\begin{eqnarray}}
\def \eeqarr{\end{eqnarray}}

\title{Response of a $d_{x^2-y^2}$ Superconductor to a Zeeman Magnetic Field}

\author{Kun Yang$^{1}$ and S. L. Sondhi$^{1,2}$}

\address{
$^{1}$Department of Physics,
Princeton University,
Princeton, NJ 08544
}
\address{
$^2$Institute for Theoretical Physics,
University of California at Santa Barbara,
Santa Barbara CA 93106-4030
}

\date{\today}
\maketitle
\begin{abstract}
We study the response of a two dimensional $d_{x^2-y^2}$
superconductor to a magnetic field that couples only to the spins of the
electrons. In contrast to the $s$-wave case, the
$d_{x^2-y^2}$ state is modified even at small magnetic fields, with the
gap nodes widening into normal, spin polarized, pockets.
We discuss the promising prospects for observing this in the cuprate
superconductors in fields parallel to the Cu-O planes.
We also discuss the phase diagram, inclusive of
a finite momentum pairing state with a novel
linkage between the momentum of the pairs and the
nodes of the relative wave function.
\end{abstract}
\pacs{}
]

Following the original work of Clogston and Chandrasekhar\cite{clogston},
the modification of superconductivity by the Zeeman coupling between the 
spins of the electrons and an applied magnetic field has attracted 
intermittent attention \cite{saintjames}. Much of this has centered on 
the bound on the upper critical field $H_{c2}$ provided by consideration of 
the Zeeman interaction 
alone (``Pauli limit'') and on the nature of the phase boundary when
this effect dominates. A more unusual aspect of this physics was uncovered 
by Fulde and Ferrell\cite{fulde} and by Larkin and Ovchinnikov\cite{larkin}
in the possibility of a finite momentum pairing state at large magnetic 
fields, where the reference Fermi surface is spin-split. Experimentally,
the classic predictions on the nature of the $s$-wave phase boundary have 
been borne out by work on thin Al films \cite{wu}, while recent work on 
unconventional
superconductors has exhibited Pauli limited critical fields \cite{norman} 
as well as the first evidence for a realization of the finite momentum 
pairing state \cite{gloos}.
Very recently, the Zeeman suppression has also been discussed in experiments
on mesoscopic samples \cite{braun}.

In this paper, motivated by the recent experimental
identification of the pairing state in several of the cuprate superconductors,
we discuss the Zeeman response of 
an ideal two-dimensional $d_{x^2-y^2}$ superconductor. We believe that this 
is a useful exercise on three grounds. First, the cuprates are strongly
two-dimensional systems and hence admit a geometry for measurements
in a magnetic field, with the field parallel to the Cu-O planes, where
the Zeeman response should dominate the orbital response at low temperatures.
Second, we estimate that spin-orbit scattering, which attenuates the 
Zeeman response, is small enough in the cleanest samples
to allow its neglect above fields as small as a few tesla.
Third and most interestingly, the existence of nodes in the gap function
imply that (in contrast to the $s$-wave case) the superconducting state 
responds non-trivially at arbitrarily small values of the magnetic field. 
As is intuitively plausible, the response is paramagnetic with the
destruction of superconductivity over parts of the Fermi surface where the
Zeeman energy $\mu B$ exceeds the local gap $\Delta({\bf k})$, and a spin
polarization of the resulting normal electrons. This leads to observable
effects in all quantities that are sensitive to the density of states
for quasi-particles.

In the following we will explicitly illustrate this by calculations on a
weak coupling BCS model and present estimates that indicate that it 
is readily observable in the cuprates. For completeness, we also discuss 
the mean-field phase diagram of our model in the field-temperature plane,
where we note the  existence of finite momentum pairing with a novel
linkage of the wave vector to the nodal structure of the gap function.

\noindent
{\bf Choice of Hamiltonian:}
We study a two-dimensional (2D) electron system described by the
Hamiltonian
\begin{equation}
H=\sum_{{\bf k}, \sigma=\uparrow, \downarrow}
\epsilon_k c^{\dagger}_{{\bf k}\sigma}
c_{{\bf k}\sigma}
+\sum_{{\bf k}, {\bf k'}}V_{{\bf k}, {\bf k'}}
c^{\dagger}_{{\bf k},\uparrow}c^{\dagger}_{{\bf -k},\downarrow}
c_{{\bf -k'},\downarrow}c_{{\bf k'},\uparrow},
\label{ham}
\end{equation}
where $\epsilon_k$ is the rotationally invariant kinetic energy 
measured from the Fermi energy $\epsilon_F$, and for $|\epsilon_k|,
|\epsilon_k'|<\epsilon_c
\ll \epsilon_F$, the pairing potential $V$ takes the form\cite{won}
\begin{equation}
V_{{\bf k}, {\bf k'}}=-2V_0\cos(2\theta_k)\cos(2\theta_{k'}) + \mu^*,
\label{V}
\end{equation}
where $V_0>0$, $\mu^*$ is the renormalized $s$-wave repulsion and 
$\theta_k$ is the azimuthal angle of ${\bf k}$; $V=0$ otherwise.
At low temperatures the system is a 
superconductor with a gap function of the $d_{x^2-y^2}$ form:
$\Delta(\theta)=\Delta_0(T)\cos(2\theta)$,
where $\Delta_0$ satisfies 
\begin{equation}
1={N(0)V_0\over 2\pi}\int_0^{2\pi}{d\theta}\int_0^{\epsilon_c}{d\xi}
{2\cos^2(2\theta)\over E(\xi, \theta)}\tanh({E(\xi, \theta)\over 2k_BT}),
\end{equation}
and $E(\xi, \theta)=\sqrt{\xi^2+\Delta_0^2\cos^2(2\theta)}$,
$N(0)$ is the single particle density of states for each spin species
at the Fermi level.
In weak coupling $N(0)V_0\ll 1$, assumed throughout this 
paper, this leads to a maximum gap $\Delta_{00}=\Delta_0(T=0)
=2.43\epsilon_c e^{-1/N(0)V_0}$, and a critical temperature 
$T_c=0.467\Delta_{00}$.
The quasi-particle spectrum is governed by the mean-field 
Hamiltonian:
\begin{equation}
H_{MF}=\sum_{\bf k}E_{\bf k}(\alpha_{\bf k}^\dagger\alpha_{\bf k}
+\beta_{\bf k}^\dagger\beta_{\bf k}),
\end{equation}
where $E_{\bf k}=\sqrt{\epsilon^2(k)+\Delta_0^2\cos^2(2\theta_k)}$, and
$\alpha_{\bf k}^\dagger$ ($\beta_{\bf k}^\dagger$) are creation
operators of up (down) spin quasi-particles. 

We should note that the $d_{x^2-y^2}$ state that we are interested in, arises 
in (nearly) tetragonal lattice systems. Our choice of model here is intended
to mimic this lattice physics with minimal and computationally
favorable ingredients: we keep a rotationally invariant Fermi surface
but introduce a pairing potential (\ref{V}) that has the 
reduced symmetry of the lattice.

A uniform magnetic field (B) applied {\em parallel} to the 2D plane does not
couple to the orbital motion of the electrons in the plane. It does, 
however, lift the spin degeneracy, and introduce the Zeeman term,
$H_Z=-\mu B \sum_{{\bf k}}(c^{\dagger}_{{\bf k}\uparrow}
c_{{\bf k}\uparrow}-c^{\dagger}_{{\bf k}\downarrow}
c_{{\bf k}\downarrow})$,
into the Hamiltonian, where $\mu=g\mu_B/2$ is the magnetic moment of 
the electron, 
and $\uparrow$
and $\downarrow$ refer to spin direction along and opposite to the field 
direction respectively. This modifies both the gap equation and the
quasi-particle Hamiltonian to 
\begin{eqnarray}
1&=&{N(0)V_0\over 2\pi}\int_0^{2\pi}{d\theta}\int_0^{\epsilon_c}{d\xi}
{\cos^2(2\theta)\over E(\xi, \theta)}\nonumber\\
&\times&[\tanh({E(\xi, \theta)+\mu B\over 2k_BT})
+\tanh({E(\xi, \theta)-\mu B\over 2k_BT})] \nonumber \\
&H_{MF}&=\sum_{\bf k}[(E_{\bf k}-\mu B)\alpha_{\bf k}^\dagger\alpha_{\bf k}
+(E_{\bf k}+\mu B)\beta_{\bf k}^\dagger\beta_{\bf k})] \ .
\end{eqnarray}
Qualitatively, the Zeeman field lowers/increases the energy 
of the spin up/down quasiparticle states which in turn changes their
occupation and affects the self-consistency condition for the gap function
(which is now distinct from the true quasiparticle gap).

\noindent
{\bf Weak-field Response:} We begin by considering weak magnetic 
fields and low temperatures: $\mu B, k_BT \ll \Delta_{00}$. 
An $s$-wave state is essentially unaffected in this limit. This is because
the occupation numbers for quasiparticle states remain
exponentially small at low $T$ and $B$, due to the finite gap,
even though the field shifts the quasiparticle bands linearly 
(at $T=0$ the gap function and the ground state are completely unaffected).

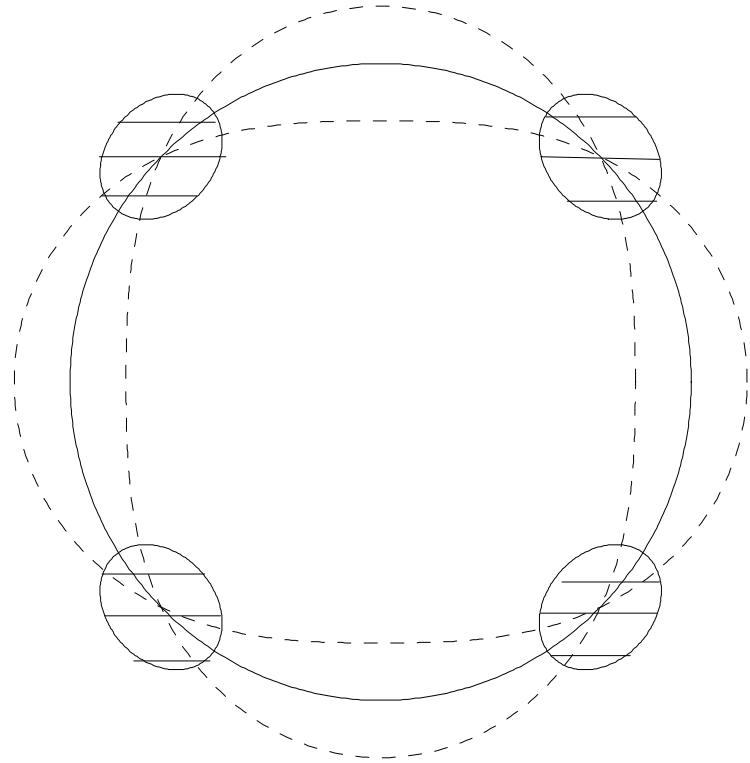
\begin{figure}
\caption{ 
Fermi surface pockets (shaded regions) produced by a Zeeman magnetic
field. The electrons in the pockets are unpaired and spin polarized along 
the direction of the field. The dashed lines indicate the extent of the 
smearing of the Fermi surface by the superconducting order at zero field, 
and show that the lateral extrema of the ``normal'' pockets are bracketed
by regions of paired electrons.
} 

\label{fig1}
\end{figure}

The situation, however, is qualitatively different in the case of the
$d_{x^2-y^2}$ state studied here; in our case the gap vanishes
at four nodal points on the Fermi surface; sufficiently close
to these points there are always some
spin up quasiparticle states whose energies become {\em negative} for
arbitrarily small values of $B$. These states, which live in elliptical
pockets, $E_k < \mu B$, near the Fermi surface (Fig. \ref{fig1}), 
develop a thermal
occupation that is $O(1)$ at {\it any} temperature. Translating back
into electron operators, one sees that these pockets are in fact regions
of the spin polarized normal state (fully polarized at $T=0$)---their inner 
and outer arcs are pieces of the spin split Fermi surface which come 
together when the angle dependent gap function exceeds the Zeeman energy. 
The loss of pairing in them and the area within $k_BT$,
leads to an overall reduction of the gap function:
\begin{eqnarray}
\Delta_0(T, B)=\Delta_{00}[1-({k_BT\over \Delta_{00}})^3
F_{\Delta}({\mu B \over k_BT}) +O(\Delta_{00}^{-4})] \nonumber \\
F_{\Delta}(x)=\int_0^{\infty}t^2[1-(\tanh{t+x\over 2}+\tanh{t-x\over 2})/2]dt\ .
\end{eqnarray}
Noting that $F_{\Delta}(0) \approx 3.61$ recovers the zero field answer, while
for $x \gg 1$ $F_{\Delta}(x) \sim |x|^3/3$ whence,
\begin{equation}
\Delta_0(T=0, B)=\Delta_{00}
[1-{1\over3}({\mu |B| \over \Delta_{00}})^3+O(B^4)]\ .
\end{equation}
Consequently, the reduction of the gap function at low fields 
and temperatures is
quite modest, and the most important effect of the proliferation of 
quasi-particles in the ground state is the {\em finite} density of 
states (DOS) at the Fermi level, which qualitatively alters the low 
energy physics of the system\cite{bahlouli}. 
We now turn to the consequences for some physical quantities in this
regime, where we may neglect the reduction of $\Delta_{0}$ at leading 
order.

\noindent {\em Specific Heat}: This takes the scaling form
\begin{eqnarray}
C(T, B)&=&2k_BN(0) {(k_BT)^2 \over \Delta_{00}} F_C({\mu B \over k_BT})
\\ \nonumber 
F_C(x)&=&\sum_{\sigma=\pm 1}
\int_0^\infty t(t+\sigma x)^2e^{t+\sigma x}/(e^{t+\sigma x}+1)^2dt\ .
\end{eqnarray}
For fields in excess of the temperature, $x\gg1$, $F_C(x)\sim {\pi^2\over 3} x$
whence $C={2\pi^2\over 3}k_B^2TN(0)\mu B/\Delta_{00}=C_N\mu B/\Delta_{00}$,
where $C_N$ is the normal state specific heat; the linear $T$ dependence at low $T$
is a consequence of the finite DOS. 
For $\mu B \ll k_BT$, we recover the expected
$T^2$ dependence upon using $F_C(0)=9\zeta(3)\approx 10.8$.
 
\noindent {\em Thermal Conductivity}: 
At low temperatures where impurity scattering may lead to a constant 
quasiparticle scattering rate\cite{lee}, 
the thermal conductivity $\kappa_e$ should be
proportional to $C$. Therefore $\kappa_e$ should increase
{\em linearly} with $T$, for $k_BT\ll \mu B$, while it  
increases with $T$ quadratically in the absence of the field, as observed 
experimentally\cite{kk}.
This is to be contrasted with the recent experimental finding\cite{kk}
that a magnetic field applied {\em perpendicular} to the Cu-O plane of
the cuprate superconductor {\em suppresses} the electronic thermal 
conductivity, presumably due to orbital effects.

\noindent {\em Magnetization:}
$M$ also takes a scaling form:
\begin{eqnarray}
M(T, B)=2\mu N(0) {(k_BT)^2 \over \Delta_{00}} F_M({\mu B \over k_BT})
\nonumber \\
F_M(x)=\int_0^\infty t({1\over e^{t-x}+1}-{1\over e^{t+x}+1})dt.
\end{eqnarray}
The limits $F_M(x\gg 1)\sim x^2/2$, and $F_M(x\ll 1)\sim (2\log 2)x$,
imply $M\propto B^2$ when $\mu B\gg k_BT$, and $M\propto B$ when $\mu B\ll
k_BT$.

\noindent {\em Tunneling Conductance}: The $T=0$ tunneling conductance of a 
superconductor-insulator-metal junction at varying bias $G(V)$ is, in 
principle, the most direct measurement of the single particle 
DOS of the superconductor. $G$ goes to zero linearly with $V$ for a
$d_{x^2-y^2}$ superconductor in zero field. For $\mu B \gg k_BT$, 
the finite DOS leads to a finite conductance: $G(V=0)=G_n\mu B/\Delta_{00}$,
where $G_n$ is the tunneling conductance when the superconductor is in its
normal state. For $eV < \mu B$, the tunneling current is spin-polarized
\cite{aleiner}.

\noindent
{\bf Phase Diagram:} We now turn to the effects of strong magnetic fields.
An $s$-wave superconductor undergoes a first order phase transition to the 
normal state when $\mu B=\Delta/\sqrt{2}$ ($\Delta$ is the $s$-wave 
gap) at $T=0$\cite{clogston}, ignoring the finite momentum pairing 
instability (see below).
This follows upon noting that the singlet $s$-wave state is
insensitive to the Zeeman field, while the normal 
state lowers its energy in proportion to its Pauli susceptibility.
The temperature-magnetic field phase diagram\cite{saintjames} exhibits a 
tricritical point where the first order line terminates, and the field tuned 
transition becomes continuous. 

The $d_{x^2-y^2}$ superconductor {\em does} respond to the Zeeman field 
but far more weakly than the normal state. This leads to a phase diagram
(Fig. 2) of the same topology as in the $s$-wave case. We find that at $T=0$ 
there
is a first order transition to the normal state at $\mu B=0.56 \Delta_{00}$
(close to the value $0.5\Delta_{00}$ obtained without accounting for the 
paramagnetism of the $d_{x^2-y^2}$ state);
at the transition the gap function has amplitude $0.92\Delta_{00}$. 
For $T<0.56 T_c$, the transition remains first order, and the normal 
(superconducting) state becomes local minimum of free energy at lower (higher)
temperature, as represented by the dotted lines in Fig. 2;
above $T=0.56 T_c$ the transition becomes continuous.

Thus far we have only considered zero momentum pairing.
At high fields one may suspect that it
might be more favorable to try to pair across the spin-split Fermi
surface of the partially polarized normal state; indeed it
is known in the $s$-wave case\cite{fulde,larkin,murthy} that this
happens at low temperatures. The pairing is then at a {\it finite} 
center of mass momentum for the Cooper pairs and the resulting 
transition from the high field normal state is continuous. At lower fields this 
Fulde-Ferrell-Larkin-Ovchinnikov (FFLO) state gives way to the zero momentum
pairing state by a first order transition. 

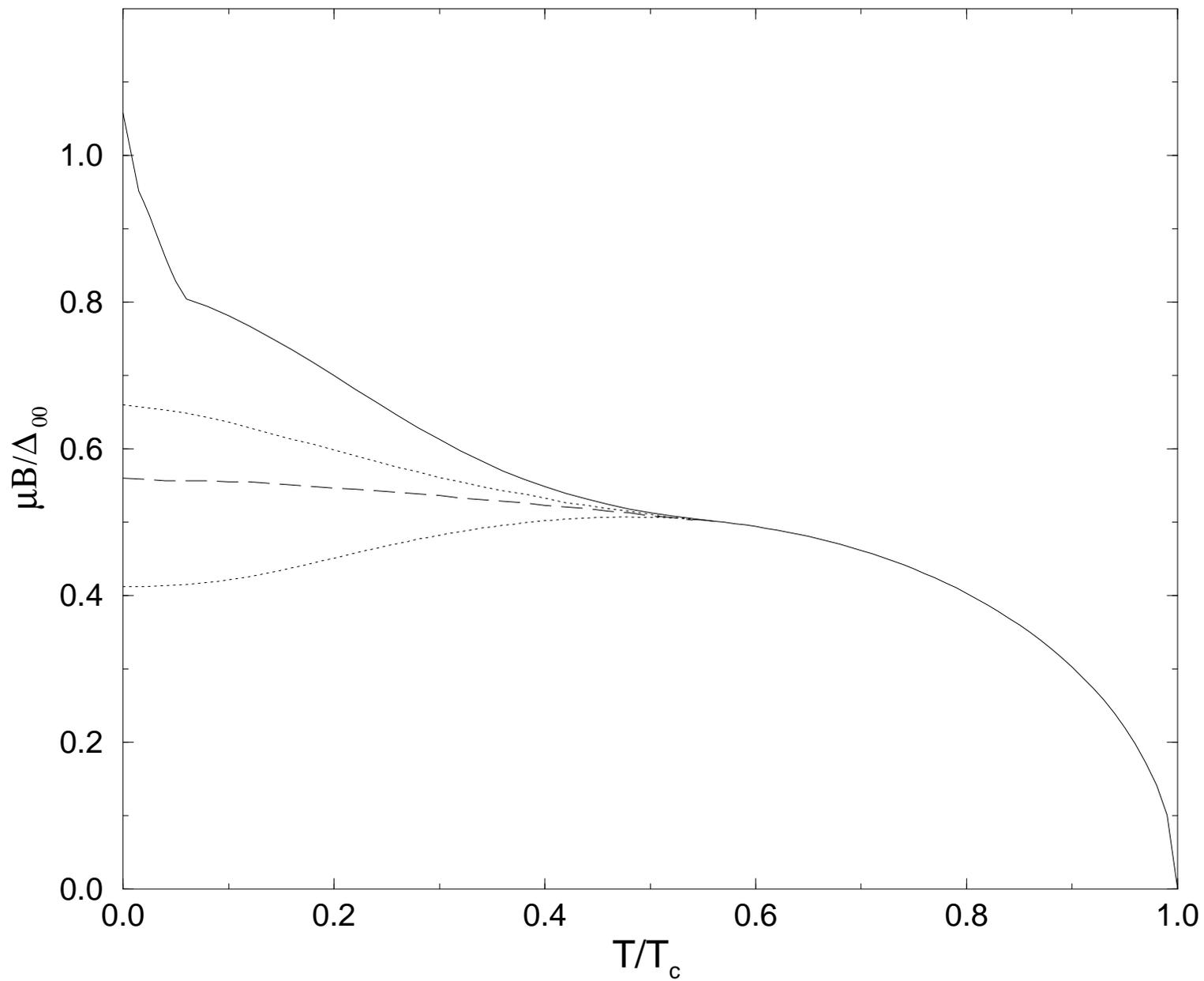
\begin{figure}
\caption{ 
The temperature-magnetic field  
phase diagram for a $d_{x^2-y^2}$ superconductor.
The solid line is the second order phase boundary separating the normal 
state and the superconducting state. 
Above $T/T_c=0.56$, the superconducting
state is the 
zero momentum pairing state while below it is a finite momentum pairing
state at high fields, which gives way to the zero momentum pairing state 
by a first
order transition along the dashed phase boundary. 
At $T/T_c\approx 0.06$, the
direction of the pairing momentum at the phase boundary changes discontinuously
from that of the gap maxima to gap minima; the kink in the phase boundary 
reflects this change.
The lower (upper) dotted 
lines are metastability lines above (below) which the 
normal (zero momentum pairing) states become local minima of the
free energy. 
} 
\label{fig2}
\end{figure}

In order to consider this possibility for the $d_{x^2-y^2}$ problem,
we extend the pairing potential in (\ref{ham}) to 
allow pairing between electrons with total momentum ${\bf q}$, 
for $q\ll\epsilon_c/v_F$,
where $v_F$ is the Fermi velocity:
\begin{equation}
\hat{V}=\sum_{{\bf k}, {\bf k'}, {\bf q}}V_{{\bf k}, {\bf k'}}
c^{\dagger}_{{\bf k},\uparrow}
c^{\dagger}_{{\bf -k}+{\bf q},\downarrow}
c_{{\bf -k'}+{\bf q},\downarrow}c_{{\bf k'},\uparrow}.
\end{equation}
As we will see later, the pairing momentum $q$ is at most of order 
$\Delta_{00}/v_F$, which is much less than $\epsilon_c/v_F$. Therefore it is 
a reasonable approximation to neglect the dependence of the pairing matrix 
element $V_{{\bf k}, {\bf k'}}$ on ${\bf q}$.
The gap still takes the $d_{x^2-y^2}$ form  and obeys
\begin{eqnarray}
1&=&{N(0)V_0\over 2\pi}\int_0^{2\pi}{d\theta}\int_0^{\epsilon_c}{d\xi}\,
{\cos^2(2\theta)\over E(\xi, \theta)}\nonumber\\
&\times&[\tanh({E(\xi, \theta)+z(\theta)\over 2k_BT})
+\tanh({E(\xi, \theta)-z(\theta)\over 2k_BT})],
\label{selfcon3}
\end{eqnarray}
where $z(\theta)=\mu B+(v_Fq/2)\cos(\theta-\theta_{\bf q})$, and 
$\theta_{\bf q}$ is the
polar angle of the pairing momentum ${\bf q}$.
In order to determine the (second order) phase boundary between the
normal and FFLO states, one needs to find the solution of (\ref{selfcon3})
with $\Delta=0$ and the largest possible $B$.
At $T=0$, we find the initial pairing instability occurs at $q=2\mu B/v_F$, 
$\theta_{\bf q}
=0, \pi$ or $\pm \pi/2$ (i.e., for ${\bf q}$ along
directions of {\em maximum} gap), 
and $\mu B\approx 1.06 \Delta_{00}$.
At lower fields the system undergoes a first order transition to the $q=0$
state. As the condensation energy of the FFLO state is quite
small (the state is gapless over much of the Fermi surface), 
an excellent estimate
for this field is simply the value 
obtained earlier for the level crossing between
the normal state and the $q=0$ state, which leads to the window
$0.56\Delta_{00}\alt \mu B \le 1.06\Delta_{00}$ for the stability of the
FFLO state, which is considerably larger than the window $\Delta/\sqrt{2}
\alt \mu B \le \Delta$ for the 
2D $s$-wave case \cite{murthy}. 
At finite temperatures, we find for $T/T_c < 0.06$, the direction of 
pairing momentum ${\bf q}$ remains the same as that of $T=0$; however
at $T/T_c > 0.06$ 
this changes {\em discontinuously} to $\theta_{\bf q}=\pm\pi/4$
or $\pm 3\pi/4$, i.e., ${\bf q}$ now points in the directions of
{\em minimum} gap. At finite $T$ we again use 
the crossing point of the free energies of the normal and $q=0$ paring states
to estimate the boundary between the zero and finite momentum pairing states
(Fig.~\ref{fig2}), this window narrows and the magnitude of the 
pairing wavevector for the
high field instability decreases and approaches
zero at $T=0.56T_c$, where the high field phase boundary 
and the coexistence line
between the $q=0$ state and the normal state come together.
In the FFLO phase, there is presumably a first order phase boundary across
which the direction of the pairing momentum changes, which ends at the 
phase boundary separating the normal and FFLO states at $T/T_c\approx 0.06$.
In the present work we have not attempted to study this phase boundary.
For $T>0.56T_c$, there is no region with $q \ne 0$ pairing and there is a
continuous transition directly from the normal state to the $q=0$ state.
This topology is also identical with that in the $s$-wave problem.
We note that our results on the phase boundary separating the normal state 
and FFLO state agree with a previous study by Maki and Won\cite{maki2},
which are also confirmed in more recent work (Ref. \onlinecite{shimahara}); 
however in these works no estimate was given for the phase boundary separating 
the FFLO state and the usual zero momentum pairing BCS state.

\noindent
{\bf Application to the cuprates:} Our analysis has been purely two
dimensional, and for such systems a magnetic field parallel to the
plane would behave precisely as a Zeeman field. For layered systems
the situation, analyzed in some detail by Klemm
{\em et al.}\cite{klb},
is more complicated for orbital effects become important near $T_c$ 
where the interplanar coherence
length is large. However, in the same approximation, there exists
a lower temperature $T^*$, where the vortex cores fit between planes
and the orbital $H_{c2}$ {\it diverges}. Below $T^*$ 
(estimated as $0.84 T_c$ and $0.99 T_c$ for YBCO and BSCCO respectively
\cite{tinkham}), assuming two dimensionality should be an excellent
approximation and hence the Zeeman response should dominate.
Another limitation of our analysis is the neglect of scalar impurity 
scattering which destroys the FFLO state in dirty superconductors, and
of spin-orbit scattering which attenuates the pair-breaking
effect of the Zeeman field. On these fronts, the news
seems to be good: the state of the art YBCO and BSCCO samples are in
the clean limit, and their residual resistivities translate 
into scattering times of order $\tau\sim10^{-12}s$, which
lead via the Elliott estimate \cite{elliott}, to a spin-orbit scattering time
$\tau_{SO} \sim \tau/(\Delta g)^2\sim 10^{-10}s$ ($\Delta g=g-2\approx0.1$).
Consequently, $\tau_{SO} > \hbar/(g \mu_B B)$ for fields above a tesla
and the neglect of spin-orbit scattering should not be too serious. A
final caveat is the conventional BCS weak-coupling nature of our analysis,
which is evidently problematic in the cuprates; absent a solution of
the larger problems in the field, we are unable to say very much more on
the issue.

Nevertheless, the qualitative physics uncovered by our model calculation,
should be quite robust to any mechanism that yields a $d_{x^2-y^2}$ state 
in a single layer \cite{fninter}. Experimentally, the low field effects 
discussed here should be readily 
observable, e.g. we estimate an enhanced specific heat 
of magnitude $0.1 HT$ mJ/mol$-K^2$ ($k_BT < \mu B$) from the existing data 
on YBCO \cite{moler}, while the high field phase transitions and the FFLO 
phase would appear only at fields of order 100T, which are currently out of 
reach. Finally, while
we have concentrated entirely on the parallel geometry, it is clear that
a full account of the response at arbitrary orientations of the magnetic
field will need to take account of the Zeeman physics discussed here.

We are grateful to A. J. Leggett,
P. W. Anderson, 
D. A. Huse, K. Krishana, K. A. Moler, N.-P. Ong, R. Shankar, 
S. A. Kivelson and D. J. Scalapino for useful 
discussions.
This work was supported in part by NSF grants DMR-9400362 (KY), 
DMR-9632690, PHY94-07194 and the A. P. Sloan Foundation (SLS).

\begin{references}

\bibitem{clogston} A. M. Clogston, Phys. Rev. Lett. {\bf 3}, 266 (1962);
B. S. Chandrasekhar, Appl. Phys. Lett. {\bf 1}, 7 (1962).

\bibitem{saintjames}  For reviews, see D. Saint-James, G. Sarma
and  E. J. Thomas, {\it Type II superconductivity}, Oxford, Pergamon Press 
(1969); P. Fulde, Adv. Phys. {\bf 22}, 667 (1973).

\bibitem{fulde} P. Fulde and R. A. Ferrell, Phys. Rev. {\bf 135}, A550 
(1964).

\bibitem{larkin} A. I. Larkin and Yu. N. Ovchinnikov, Sov. Phys.-JETP
{\bf 20}, 762 (1965).

\bibitem{wu} W. Wu and P. Adams, Phys. Rev. Lett. {\bf 73},
1412 (1994).

\bibitem{norman} 
R. H. Heffner and M. R. Norman, Comments on Condensed Matter Physics, 
{\bf 17}, 361 (1996).

\bibitem{gloos} K. Gloos {\it et. al.}, Phys. Rev. Lett {\bf 70}, 501 (1993).

\bibitem{braun} F. Braun {\it et. al.}, 
Phys. Rev. Lett {\bf 79}, 921 (1997).

\bibitem{won} H. Won and K. Maki, Phys. Rev. B {\bf 49}, 1397 (1994).

\bibitem{murthy} L. N. Bulaevskii, Sov. Phys.-JETP {\bf 38}, 634 (1974);
H. Shimahara, Phys. Rev. B {\bf 50}, 12760 (1994);
G. Murthy and R. Shankar, J. Phys. Cond. Matt.  {\bf 7}, 9155 (1995).

\bibitem{bahlouli} A similar effect was discussed in the context of three
dimensional heavy fermions by H. Bahlouli, Phys. Lett. A {\bf 118}, 209
(1986).

\bibitem{lee} We note that while the data in Ref. \onlinecite{kk} 
supports the idea of a constant scattering rate, there is work 
suggesting otherwise: P. A. Lee, Phys. Rev. Lett {\bf 71},
1887 (1993); L. Taillefer {\em et al.}, Phys. Rev. Lett {\bf 79}, 483 (1997).

\bibitem{kk} K. Krishana {\em et al.},
Science {\bf 277}, 83 (1997).

\bibitem{aleiner} Even in this regime, superconducting fluctuations are
likely to lead to an anomalous tunneling conductance along the lines
discussed by I. L. Aleiner and B. L. Altshuler, Phys. Rev. Lett.
{\bf 79}, 4242 (1997).

\bibitem{maki2} K. Maki and H. Won, Czech. J. Phys. {\bf 46} S2, 1035
(1996).

\bibitem{shimahara} H. Shimahara and D. Rainer, J. Phys. Soc. Jpn.
{\bf 66}, 3591 (1997); H. Shimahara, preprint cond-mat/9711017 (1997).

\bibitem{klb} R. A. Klemm {\em et al.}, Phys. Rev. B {\bf
12}, 877 (1975).

\bibitem{tinkham} M. Tinkham, {\em Introduction to Superconductivity},
2nd edition, McGraw Hill, New York, 1996.

\bibitem{elliott} S. R. Elliott, Phys. Rev. {\bf 96}, 266 (1954).

\bibitem{fninter} The inter-layer tunneling model of Anderson
and co-workers, see e.g. P. W. Anderson, J. Phys. Chem. Solids {\bf 56},
1593 (1995), is fundamentally different in this regard.

\bibitem{moler} K. A. Moler {\it et. al.}, Phys. Rev. B {\bf 55}, 3954 (1997).

\end{references}
\end{document}